# Purcell-enhanced Bright and Dark Exciton Emission from Perovskite Quantum Dots in Micro-ring Resonators

*Lanyin Luo[1,4], Mohit Khurana[1,4], Ian M. Murray[2], Sina Baghbani Kordmahale[1,5], Akanksha Pandey[1], Xiaohan Liu[1], Alexei V. Sokolov[1,4], Dong Hee Son[1,2,3]\**

[1]Department of Physics and Astronomy, Texas A&M University, College Station, Texas, 77843, USA.

[2]Department of Chemistry, Texas A&M University, College Station, Texas, 77843, USA.

[3]Center for Nanomedicine, Institute for Basic Science and Graduate Program of Nano Biomedical Engineering, Advanced Science Institute, Yonsei University, Seoul 03722, Republic of Korea.

[4]Institute for Quantum Science and Engineering, Texas A&M University, College Station, TX 77843.

[5]Department of Electrical and Computer Engineering, University of Massachusetts Amherst, Amherst, MA 01003

**ABSTRACT**

Colloidal quantum dots (QDs) integrated with waveguide-coupled dielectric resonators are promising building blocks for compact on-chip light sources. However, deterministic placement of QDs with strong mode overlap at the desired location remains a challenge. Here, we demonstrate a simple and scalable strategy for integrating colloidal QDs with a waveguide-coupled $Si_3N_4$ micro-ring resonator platform and for controlling the radiative dynamics of both bright and dark excitons via Purcell enhancement. We use strongly quantum-confined $CsPbBr_3$ QDs, which exhibit bright-exciton emission at room-temperature, while emission at cryogenic temperatures originates from both bright and dark excitons. The $CsPbBr_3$ QDs are selectively retained on the $Si_3N_4$ micro-ring cavities through a spin-coating/rinsing process, enabling efficient overlap with whispering-gallery modes and routing of the emission through integrated waveguides. We confirm accelerated decay of emission from both bright and dark excitons for $CsPbBr_3$ QDs coupled to the micro-ring cavities. These results demonstrate an effective route to integrate colloidal QDs with $Si_3N_4$ micro-ring cavities and to leverage cavity-enhanced emission in scalable integrated photonic devices.

**INTRODUCTION**

Integrated photonic devices that incorporate colloidal quantum dots (QDs) provide a versatile route to engineer light-matter interactions at the micro- and nanoscale, leveraging the strong emission and solution processability of colloidal QDs. With widely tunable bandgaps, high photoluminescence (PL) quantum yields, and compatibility with low-temperature fabrication, colloidal QDs are attractive building blocks for on-chip photonics. Accordingly, resonant cavities integrated with colloidal QDs have been explored for various applications including single-photon emission, spontaneous emission control and lasing.[1-3]

Among cavity geometries, micro-ring resonators are especially well suited for Purcell enhancement of emission in integrated photonics because they combine high-Q whispering gallery modes with deterministic coupling to well-defined waveguides for efficient photon collection and routing. While micro-disks also support high-Q whispering gallery modes, micro-rings enable more controllable in/out coupling through a bus waveguide and facilitate device-level integration into larger photonic circuits.[4-6] Compared with Fabry-Perot cavities, which rely on two reflective interfaces to establish resonances, micro-rings offer compact footprints, straightforward on-chip waveguide integration, and intrinsically narrowband wavelength selectivity.[7] In addition, the resonance spectrum set by the radius and cross-section of the ring provides a tunable free spectral range (FSR), enabling precise spectral alignment between cavity modes and QD emission to strengthen emitter-cavity coupling.[8-12]

Despite these advantages, reports of cavity-coupled optical properties of colloidal QDs in micro-ring resonators remain limited. A key bottleneck is placing the emitters at predetermined locations with sufficient spatial precision and strong overlap with the cavity field. For this reason, many early demonstrations relied on epitaxial QDs with deterministic positioning during growth.[13-17] However, epitaxial integration typically requires high-temperature processing that is often incompatible with prefabricated dielectric photonic circuits and colloidal QDs. Alternative approaches for integrating colloidal QDs, such as spin-coated films or lithographically defined placement templates, have enabled coupling to dielectric cavities, but often require post-fabrication resist processing that increases process complexity and may compromise the QD's structural or optical performance.[18-21] More recently, inkjet printing has enabled deterministic placement of individual colloidal QDs without using resist, but additional development is needed to translate these demonstrations into a broadly practical approach.[22, 23] These challenges motivate the practical and scalable methods that can achieve high field overlap while avoiding processing incompatible with colloidal QDs.

Here, we employed a simple spin coating and rinsing process that enables scalable, selective deposition of colloidal QDs onto an array of $Si_3N_4$ micro-ring resonators with integrated waveguides on a $SiO_2$/Si substrate, alleviating a key challenge in deterministic QD placement. This approach concentrates QDs on the top surface of the micro-ring cavity, maximizing overlap with the evanescent field of the whispering gallery modes of the cavity for efficient emitter-cavity interactions. We chose $CsPbBr_3$ QDs for this study, since metal halide perovskite QDs offer superb photon emission capability, although they are more susceptible to structural changes in polar solvent environment compared to other colloidal QDs.[24, 25] Moreover, in strongly quantum-confined $CsPbBr_3$ QDs, emission from the dipole-forbidden dark exciton can also be accessed at low temperatures, allowing us to explore cavity-enhanced emission from both bright and dark exciton states.[26-28] Recent studies reported the cavity-coupled emission of perovskite QDs using various cavity structures, such as circular-Bragg grating cavities, $Si_3N_4$ nanobeam cavities, microdisk cavities, and Fabry- Pérot cavity, while these efforts largely focused on cavity-coupling of bright exciton and have not addressed scalable deterministic QD integration.[3, 7, 29-32] In this study, we observed stable Purcell-enhanced emission of both bright and dark excitons of $CsPbBr_3$ QDs using the QD-incorporated micro-ring resonator arrays, with emission routed through the integrated waveguide and a grating coupler. These results demonstrate the potential of our approach as a scalable route to cavity-enhanced photon sources based on colloidal QDs on an on-chip photonic platform that enables flexible photon routing through integrated waveguides.

**RESULTS AND DISCUSSION**

To investigate Purcell enhancement of bright and dark excitons from colloidal $CsPbBr_3$ QDs integrated with a micro-ring resonator, we fabricated $Si_3N_4$ micro-ring cavities with accompanying bus waveguides and grating couplers on a $Si/SiO_2$ substrate using a standard lift-off process. The details of the fabrication process are provided in the Experimental section. Figure 1a shows a scanning electron microscope (SEM) image of a representative micro-ring cavity selected from the array of cavities fabricated on a $Si/SiO_2$ substrate. The gratings on the bus waveguides are used to couple excitation light into the micro-ring cavity and to out-couple photoluminescence (PL) from the $CsPbBr_3$ QDs deposited on the ring. The micro-ring waveguides are 300 nm wide and 200 nm tall. The minimum gap between the bus waveguide and the ring is 120 nm, enabling the coupling of light between the micro-ring and the bus waveguide necessary for both excitation of the QDs on the micro-ring cavity and detection of PL from the QDs. Additional SEM images showing the array of cavities and details of the waveguide and grating structures are provided in in Supporting Information (Figure S1) Figure 1b shows the schematic cross-sectional view of the device. The refractive index contrast between $Si_3N_4$ (~2) and $SiO_2$ (~1.45) enables total internal reflection in the resonator, effectively confining the PL from $CsPbBr_3$ QDs deposited on top of the $Si_3N_4$ micro-ring within the cavity with low loss.

Ideally, $CsPbBr_3$ QDs should be deposited only on top of the micro-rings to evaluate Purcell enhancement of the exciton emission without interference from the QDs deposited at other locations. To overcome the difficulties associated with fabricating QD-incorporated cavity structures with deterministic QD positioning, we employed a simple spin-coating/rinsing procedure that preferentially deposits $CsPbBr_3$ QDs on $Si_3N_4$ micro-ring rather than on $SiO_2$ region. This approach takes advantage of the difference in the oleophilicity of $Si_3N_4$ and $SiO_2$ surfaces,[33] such that QDs passivated with non-polar surface ligands are preferentially deposited on $Si_3N_4$ micro-ring.

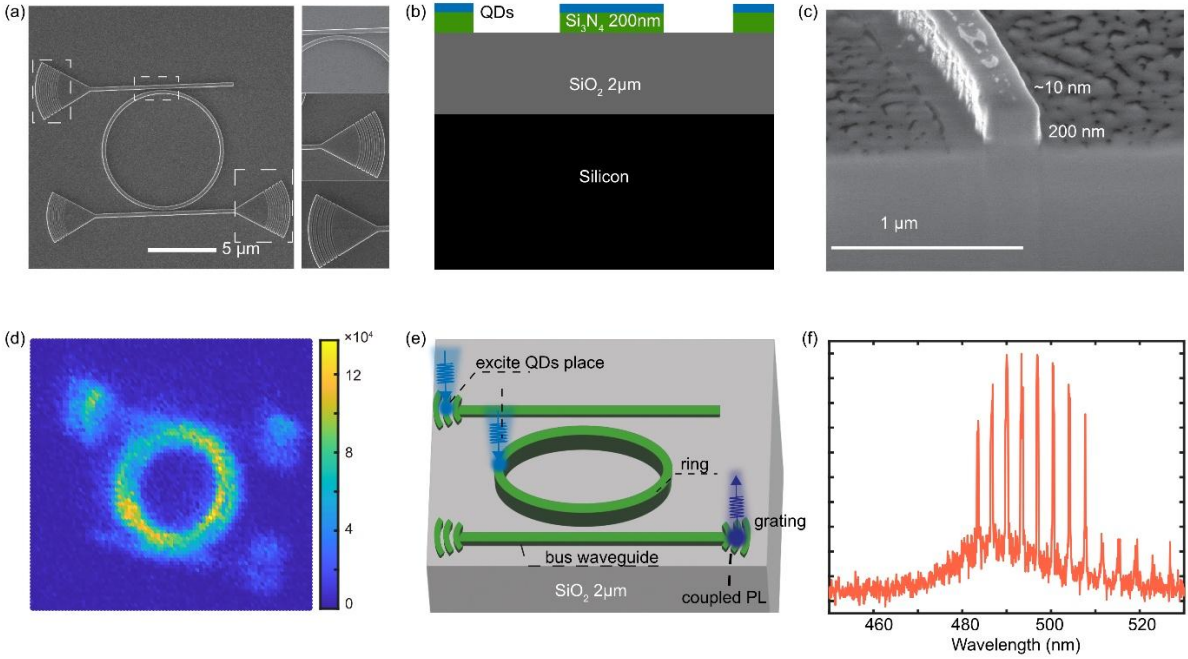

**Figure 1.** Structure of the micro-ring cavity resonator and experimental scheme. (a) Scanning electron micrograph (SEM) image of the micro-ring cavity. The magnified views of the dashed rectangle regions are shown on the right side. (b) Schematic diagram of cross-sectional view of the cavity structure. (c) SEM image showing an exposed cross-sectional view of the cavity. (d) Scanning confocal PL microscope image of the QDs integrated with cavity. (e) Locations of excitation and PL collection. (f) PL spectrum from cavity-coupled $CsPbBr_3$ QDs collected at the grating of the lower bus waveguide.

Briefly, a concentrated solution ($1.8\times10^{-5}$ M) of oleylammonium bromide-passivated $CsPbBr_3$ QDs dispersed in toluene was first spin-coated to form a continuous QD film over the entire device area. A subsequent rinsing with toluene removed loosely bound QDs from the less oleophilic $SiO_2$ surface much more easily than those on more oleophilic $Si_3N_4$ surface, thereby retaining QDs preferentially on $Si_3N_4$ surfaces. Figure 1c shows an SEM image of a micro-ring cavity after QD deposition using the described procedure with a cross-sectional region exposed by focused ion beam (FIB) milling. A thin (~5 nm) gold layer was evaporated onto the QD-deposited micro-ring device prior to imaging to minimize charging and improve resolution. The image confirms that QDs remain on the top surface of the $Si_3N_4$ micro-ring, although they do not fully cover the surface due to partial removal during rinsing. Preferential deposition of QDs on $Si_3N_4$ surface through the spin-coating/rinsing process was also corroborated by fluorescence microscopy of the device. Figure 1d shows a scanning confocal microscope image of the cavity, where PL originates predominantly from the micro-ring and bus waveguide regions, while PL from $SiO_2$ regions is more than two orders of magnitude weaker. An additional confocal fluorescence image of an array of cavities, demonstrating scalability to multiple cavity structures, is provided in Supporting Information (Figure S1).

Figure 1e illustrates the geometry of excitation and detection of the PL from the cavity-coupled QDs by using a confocal microscope that can excite and detect at different locations using two independently operating galvo mirrors. The excitation light can be delivered through the input grating on the upper bus waveguide, and the cavity-coupled PL can be collected at a grating on the lower bus waveguide. Because QDs are also present on the waveguide and grating regions, excitation through the grating can experience partial attenuation along the bus waveguide. Therefore, we also excited QDs directly on top of the micro-ring, which reduced propagation loss for the excitation light. The cavity-coupled PL from the $CsPbBr_3$ QDs, collected through the output waveguide and grating, exhibits resonance modes determined by the ring diameter. Figure 1f shows the cavity-coupled PL spectrum of $CsPbBr_3$ QDs at room-temperature. The PL spectrum of the QDs on the micro-ring shows whispering gallery modes of $Si_3N_4$ micro-ring cavity superimposed on a weak, broad background. We attribute the broad background to emission from QDs that are not efficiently coupled to the cavity mode, likely originating from QDs near the output grating coupler region. The cavity-coupled PL spectrum shows evenly spaced resonances with a free spectral range (FSR) of ~3.3 nm, which is consistent with the value obtained from simulation as discussed later. After the deposition of QDs, the cavity's quality factor ($Q$), estimated from the resonance peak position ($\lambda_0$) and the linewidth ($\Delta\lambda$) of the peak as $Q = \frac{\lambda_0}{\Delta\lambda}$, is approximately 1000. QDs deposited on the cavity devices can be removed readily by sonication in hexane without affecting the device, enabling reuse for repeated deposition and measurement cycles. The cleaning and reuse procedure is described in the Experimental section.

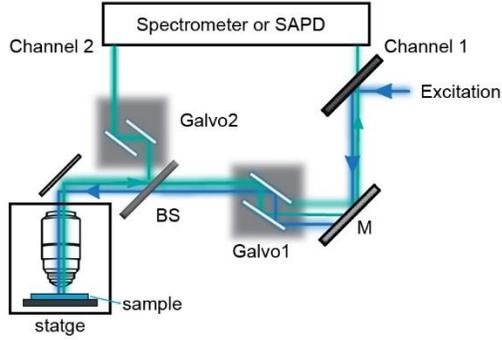

**Figure 2.** Schematic diagram of the experimental setup.

In this study, we chose strongly quantum-confined CsPbBr$_3$ QDs with an average size of ~4.2 nm, which is significantly smaller than twice the exciton Bohr radius ($2a_B \approx$ 7nm). Previous studies reported that both dipole-allowed bright exciton and dipole-forbidden dark exciton PL can be observed at low temperatures (e.g., < 20 K) for strongly confined CsPbBr$_3$ QDs, while bright exciton PL dominates at higher temperatures.[27] This makes it convenient to examine the Purcell-enhanced decay of both bright and dark excitons using the same colloidal QDs and the same device. To characterize the spectral and temporal properties of cavity-coupled exciton PL from CsPbBr$_3$ QDs, we used a confocal microscope with two independently controlled beam paths using two 2D scanning galvo mirrors as shown in Figure 2. One path (Galvo 1) was used for excitation and conventional scanning confocal imaging. The second path (Galvo 2) was used for the detection of PL at spatially offset locations from the point of excitation, enabling efficient collection of cavity-coupled emission at the output grating, as indicated in Figure 1e. The PL from the QDs was directed to either a charge-coupled device (CCD) spectrometer or a single-photon counting avalanche photodiode (SAPD) in a time-correlated single-photon counting (TCSPC) system for spectrally or temporally resolved measurements.

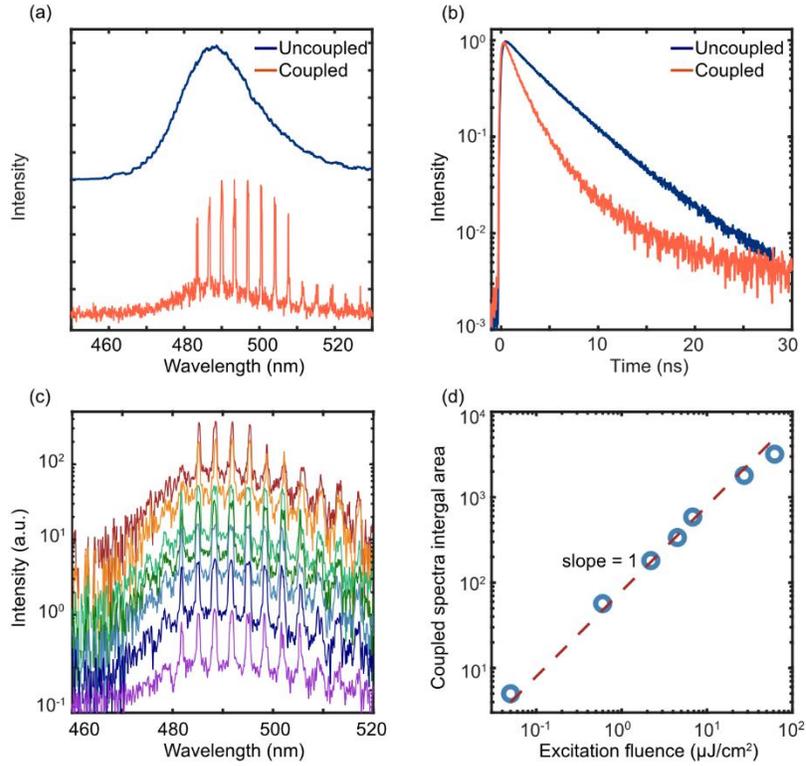

**Figure 3**. (a) Steady-state PL spectra of uncoupled QDs (blue) and cavity-coupled QDs (orange). (b) Time-resolved PL decay of uncoupled (blue) QDs and cavity-coupled QDs (orange) (c) Excitation fluence dependence of the cavity-coupled QDs spectra. Fluence (in $\mu J/cm^2$) from bottom to top: 0.05, 0.6, 2.2, 4.5, 6.8, 27.3, 62.1 (d) Spectrally integrated cavity-coupled PL intensity as a function of excitation fluence.

To compare the spectra and exciton emission lifetimes of the cavity-coupled and uncoupled $CsPbBr_3$ QDs on the same device, the signal from uncoupled $CsPbBr_3$ QDs was collected at a location far from the micro-ring cavity on $SiO_2$ region of the device. While the signal from $SiO_2$ region is two orders of magnitude weaker than that from the micro-ring region due to the low density of QDs, reliable spectra and emission lifetimes could be obtained with sufficiently long data acquisition. Figure 3a and 3b compare the steady-state spectra and time-resolved PL intensities from cavity-coupled and uncoupled $CsPbBr_3$ QDs at room-temperature. At room-temperature, exciton PL from $CsPbBr_3$ QDs originates entirely from bright exciton, while the contribution of dark exciton becomes more apparent as the temperature is lowered below 20 K.[27] Therefore, the data obtained at room-temperature provide information on Purcell enhancement of bright exciton only. Figure 3a compares the spectrum from the cavity-coupled and uncoupled QDs. To quantify the influence of the cavity on the emitter's radiative properties, the Purcell factor is estimated by comparing the lifetimes of cavity-coupled and uncoupled QDs as shown in Figure 3b. The measured lifetimes are 1.6 ns and 4.5 ns, respectively, for cavity-coupled and uncoupled QDs. This difference does not originate from differences in the substrate or the QD density on $Si_3N_4$ micro-ring and $SiO_2$, as confirmed in a separate comparative measurement. The comparison of the spectra and lifetimes of the PL from the QDs deposited on flat $Si_3N_4$ and $SiO_2$ substrates at a QD concentration comparable to that of the QDs on $Si_3N_4$ micro-ring cavity indicates that the substrate and QD density have little effect on the lifetime (Figure S2).

The Purcell factor can be expressed as $F_p = \tau_0/\tau_{cav}$, where $\tau_0$ and $\tau_{cav}$ are the radiative decay times of uncoupled and cavity-coupled QDs respectively. The PL quantum yield (η) of the QDs in solution phase used in this study was determined to be ~0.6 at room-temperature. Assuming that QDs deposited on the substrate have a similar PL quantum yield, we obtain a Purcell factor of $F_p = (\tau_u/\tau_c - 1 + \eta)/\eta \approx 4$ taking into account the finite quantum yield, where $\tau_c$ and $\tau_u$ are the measured lifetimes of the PL from cavity-coupled and uncoupled QDs respectively.[34] Measurements made on other cavities fabricated on the same substrate result in similar values.(Figures S3 and S4)

Figure 3c and 3d show the dependence of the cavity-coupled bright exciton emission spectrum and intensity on excitation fluence. The data show a linear increase in signal intensity with excitation fluence and no change in the spectrum as a function of excitation fluence, indicating the absence of multiexciton or nonlinear effects over the range of excitation fluence used (0.05-62 µJ/cm$^2$). In previous studies on dielectric whispering gallery mode cavities, Purcell factors ($F_p$) have been reported to be in the range of 1.3 to 8, typically in vertically stacked microdisk structures.[29, 31, 35] $F_p$ of ~4, achieved with a single micro-ring cavity, demonstrates a substantial radiative rate enhancement in a simpler, fully integrated platform.

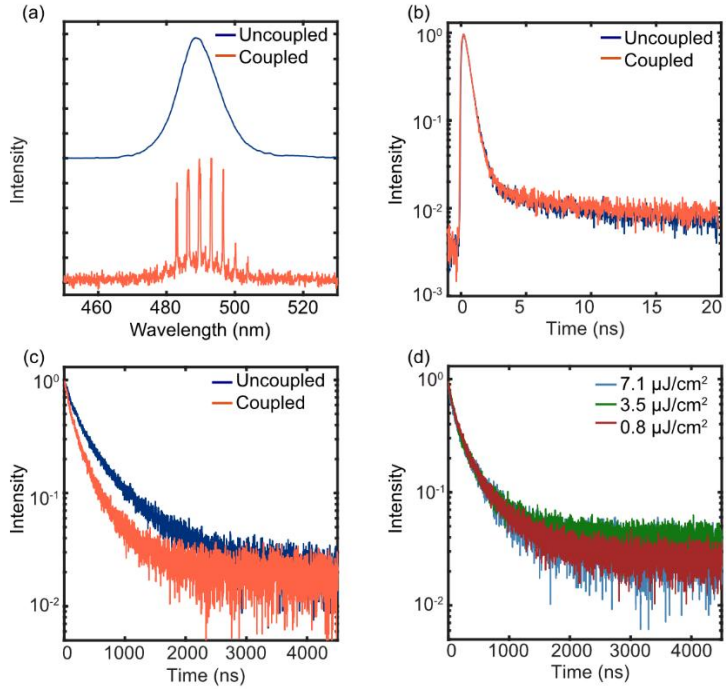

**Figure 4**. (a) Steady-state PL spectra of uncoupled QDs (blue) and cavity-coupled QDs (orange) at 10 K. (b) Time-resolved PL decay of uncoupled QDs (blue) and cavity-coupled QDs (orange) at 10 K. (c) Dark exciton decay of uncoupled QDs (blue) and cavity-coupled (orange) QDs at 10 K. (d) Excitation fluence dependence of the dark exciton decay from cavity-coupled QDs at 10 K.

According to earlier studies on strongly quantum-confined $CsPbBr_3$ QDs, the dark exciton state is the lowest energy state, with the energy gap between the bright and dark states increasing with increasing quantum confinement.[27, 28] At sufficiently low temperatures, where thermal excitation from the dark to the bright state is suppressed, the contribution of the longer-lived dark exciton to the PL becomes prominent. Consistent with prior reports, we observed dark exciton PL from 4.2 nm $CsPbBr_3$ QDs below ~20 K. Figure 4a compares the PL spectra of cavity-coupled and uncoupled $CsPbBr_3$ QDs at 10 K. The time-resolved PL intensity data are shown in Figure 4b and c in two different time windows. These data exhibit two distinct decay components: a fast component of ~460 ps and a slow component on hundreds of ns timescale. The faster component is consistent with previously reported values for similar-sized $CsPbBr_3$ QDs and reflects the combined dynamics of the bright exciton emission and decay from the bright to the dark state.[27] The slower component is attributed to dark exciton recombination. The cavity-coupled dark exciton lifetime is reduced (170 ns) relative to that of uncoupled QDs (270 ns), corresponding to an apparent enhancement of the decay rate by ~1.6, while the fast component remains nearly unchanged. The accelerated dark exciton decay remains constant as the excitation intensity is increased by nearly an order of magnitude (Figure 4d), showing that enhanced decay is free from nonlinear multi-exciton effects. Since the PL quantum yield of the QDs in the device at 10 K is uncertain in this study, we assume it to be larger than 0.6 based on the earlier studies showing generally increasing quantum yield with decreasing temperature.[27, 36] The Purcell enhancement of $F_p < 2$ for the dark exciton determined from the lifetime is smaller than that of the bright exciton at room-temperature, assuming η > 0.6 at 10 K. In contrast, the insensitivity of the fast component is more intriguing. Although a substantial fraction of the fast decay component arises from the nonradiative decay from the bright to the dark state, which is not affected by the cavity, the absence of a clear signature of acceleration is somewhat unexpected.

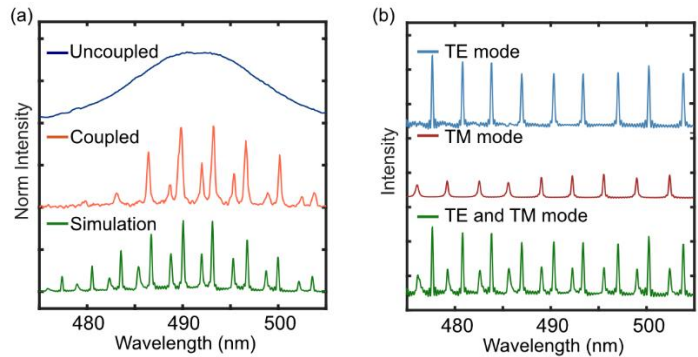

**Figure 5.** (a) Comparison of experimental PL spectra of uncoupled QDs (blue) and cavity-coupled QDs (orange) at 10 K and simulated spectrum showing both TE and TM modes (green). (b) Simulated spectra of cavity-coupled QDs with the transition dipole oriented parallel (blue) and perpendicular (red) to the plane of the substrate assuming 10 nm thick film of the QDs are on the cavity. For comparison, the spectrum showing both TE and TM modes are also included.

While the spectra from the cavity-coupled QDs exhibit predominantly a single set of resonance modes in most of the samples we examined, we also observed the spectra exhibiting two different sets of modes as shown in Figure 5a. We assigned these to fundamental transverse electric (TE) and transverse magnetic (TM) modes with different effective indices and therefore different FSR values, whereas the TE mode dominates when a single set of resonance modes is observed. This assignment is supported by the finite-difference time-domain (FDTD) simulations of TE and TM modes for the micro-ring cavity used in this study. Lumerical was used for this calculation assuming that a 10 nm-thick QD film was deposited on top of $Si_3N_4$ micro-ring cavity. Figure 5b compares the spectra of the individual calculated TE and TM modes together with the spectrum of the combined modes assuming a flat PL spectrum within the wavelength range displayed in the figure. The spectrum of the combined TE and TM modes in Figure 5b, weighted by the envelope of the uncoupled QD spectrum is shown below the experimentally measured cavity-coupled spectra in Figure 5a. The peak positions and FSR values match well between the two spectra supporting the assignment to TE and TM modes.

Fundamental TE and TM modes arise from the emitters with its transition dipoles parallel and perpendicular to the substrate respectively. Since cube-shaped $CsPbBr_3$ QDs possess transition dipoles in all three mutually orthogonal directions,[26] both modes should be visible when $CsPbBr_3$ QDs are deposited with random orientation on the cavity. In our study, simultaneous appearance of both TE and TM modes was more frequently observed when the concentration of $CsPbBr_3$ QDs used for deposition was lower ($<4\times10^{-6}$M). There are many factors that influence the intensity of each mode other than the orientation of the emitter dipole, such as the propagation loss influenced by the quality of the cavity and bend loss. Self-absorption and scattering can also contribute to the intensity of the detected resonance modes.[37-41] The fundamental TM mode has a larger out-of-plane field than TE mode, with a stronger overlap with the QD layer at the upper surface. Therefore, we conjecture that the propagation loss at the higher QD concentration is more substantial for TM mode, resulting in its suppression and observation of predominantly TE mode when the device was fabricated with high QD concentrations.

We also performed measurements in a fully waveguide-driven, grating to grating configuration, where both excitation and collection occur at the grating couplers rather than directly on the ring (Figure S5). In this geometry, the Purcell-enhanced emission is launched into the bus waveguide and micro-ring, routed through another waveguide, and extracted from a remote grating, demonstrating that cavity-modified QD emission can be generated, guided, and read out entirely within the integrated photonic circuit without direct optical access to the cavity region. Confocal PL images of the spatial propagation of the cavity-coupled QD emission for different excitation areas are provided in Figure S6.

**CONCLUSIONS**

We developed a reusable, waveguide-integrated $Si_3N_4$ micro-ring cavity platform that enables coupling of colloidal $CsPbBr_3$ QDs to whispering gallery modes using a simple spin coating/rinsing process that preferentially retains QDs on $Si_3N_4$ surfaces. Cavity coupling produces Purcell-enhanced emission from both bright and dark excitons. At room temperature, where bright exciton emission dominates, we observe a substantial enhancement of bright exciton emission. At cryogenic temperature (10 K), the long-lived dark exciton component shows Purcell-enhanced acceleration, whereas the bright exciton component does not show a clear signature of acceleration. The QD deposition method also enables facile removal and redeposition of QDs on the $Si_3N_4$ cavity device, allowing reuse and iterative optimization on the same photonic circuit. Overall, this work provides a practical and scalable route to integrate colloidal QDs with waveguide-coupled $Si_3N_4$ micro-ring cavities and demonstrate Purcell-enhancement of both bright and dark exciton emission within an integrated photonic circuit.

**EXPERIMENTAL SECTION**

**Synthesis of $CsPbBr_3$ QDs**

Strongly quantum-confined $CsPbBr_3$ QDs were synthesized following the previously published procedures.[42, 43] The Cs precursor was prepared by dissolving $Cs_2CO_3$ (0.6 g) in oleic acid (OA, 2.4 g) and octadecene (ODE, 6.4 g) on a Schlenk line under vacuum, followed by switching to a $N_2$ atmosphere at 120°C for 10 min. The Pb/Br precursor was prepared by dissolving $PbBr_2$ (300 mg) and $ZnBr_2$ (700 mg) in a mixture of OA (7 mL), oleylamine (OAm, 7 mL) and ODE (20 mL). In a flask, the Pb/Br mixture was heated at 120 °C under vacuum for 5 min. To initiate the reaction, 3.2 mL of the Cs precursor solution was injected into the Pb/Br precursor under $N_2$ and allowed to react for 1 hour at 100 °C. The reaction mixture was then cooled to below 50 °C. NaBr-saturated acetone was added to the crude solution to precipitate the QDs. After centrifugation at 6000 rpm for 5 minutes, the precipitated QDs were recovered and redispersed in hexane for further use.

**Micro-ring Cavity Fabrication**

Micro-ring cavity structures were fabricated using a standard lift-off-based patterning and etching workflow.[11, 44, 45] Briefly, the starting wafer consisted of a 200 nm low-pressure chemical vapor deposition (LPCVD) silicon nitride ($Si_3N_4$) film on 2 μm of $SiO_2$ grown on a Si substrate. Before fabrication, the wafer was cleaned with an acetone/isopropanol (IPA) mixture and dried on a hot plate. Poly(methyl methacrylate) (PMMA) 950 A4 was spin-coated at 4000 rpm for 1 minute and soft-baked at 180 °C for 2 minutes. Micro-ring patterns were written using a TESCAN electron-beam lithography (EBL) system with an electron dose of 450 μC cm$^{-2}$, a beam current of 0.260 nA, and an accelerating voltage of 30 kV. The patterns were developed in MIBK:IPA (1:3) for 1 minute and dried with nitrogen. A 30 nm chromium (Cr) hard mask was deposited by electron-beam evaporation. The wafer was immersed vertically in acetone for more than 2 hours to remove PMMA. Reactive Ion Etching (RIE) was performed with 200 W RF forward power using 5 standard cubic centimeter per minute (sccm) $O_2$ and 50 sccm $CHF_3$ for 7 min to etch the $Si_3N_4$ layer. After RIE, the remaining Cr mask was removed in Cr etchant for 5 minutes. The wafer was rinsed with deionized water and dried. The detailed process flow is shown in Figure S7.

**QD Deposition and Device Cleaning Procedure**

For device preparation, a concentrated solution of CsPbBr$_3$ QDs in hexane ($1.8\times10^{-5}$ M) was used to spin-coat the device at 3000 rpm. Subsequently, 10 μL of hexane was dropped vertically onto the surface to rinse the chip, and the substrate was dried with a nitrogen gun. This procedure promotes preferential retention of QDs on the Si$_3$N$_4$ photonic structures relative to the SiO$_2$ regions. The QD film was stripped by sonication in hexane for 10 minutes in a beaker, while keeping the cavity-facing surface of the device from contacting the beaker to avoid mechanical abrasion. The chip was then dried with a nitrogen gun and reused for the next deposition cycle.

**Optical Measurements**

Solution-phase absorption spectra of CsPbBr$_3$ QDs were obtained using a CCD spectrometer (Ocean Optics, USB2000) equipped with a deuterium light source (UV-VIS ISS, Ocean Optics). Steady-state PL spectra of QD solutions were collected using the same spectrometer under 365 nm UV-lamp excitation. Steady-state PL spectra of the QDs on the micro-ring cavity devices were measured using a home-built confocal microscope equipped with a dual-grating spectrograph (Princeton Instruments, SP-2150i) and a CCD camera (Princeton Instruments, PIXIS 100). Bright exciton lifetimes were measured using a 405 nm diode laser (PicoQuant, P-C-405; 45 ps pulse width; 5 MHz repetition rate). Dark exciton lifetimes were measured using a 405 nm continuous-wave laser (Thorlabs, L405G2) modulated with an acousto-optic modulator and a digital delay generator (Stanford Research, DG645) to generate 100 ns pulses at a 10 kHz repetition rate. For cryogenic measurements, the objective was integrated into a closed-cycle cryostat system (Janis ST-500). The confocal setup employed two independently addressable 2D galvo mirror pair (Thorlabs GVS212) to separate the excitation spot from the collection position with a 65 μm × 65 μm field of view. Channel 1 and Channel 2 collected PL signals through a 50:50 beam splitter. For propagation mapping and grating readout, Galvo 1 was held fixed at the excitation location while Galvo 2 raster-scanned the collection spot to sample PL at spatially offset positions, including the output grating for cavity-coupled emission. PL lifetimes were acquired by sending the avalanche photodiode (APD) signal to a TCSPC module (PicoHarp 330), and spectra were collected via an optical fiber coupled to the CCD spectrometer.

## Data availability

All data supporting this article are included in the main text and the Supporting Information. Raw data are available from the corresponding author upon reasonable request.

## AUTHOR INFORMATION

### Corresponding Author

*Corresponding author Dong Hee Son. Department of Chemistry, Texas A&M University, College Station, Texas 77843, USA. E-mail: dhson@chem.tamu.edu

### Author Contributions

All authors have given approval to the final version of the manuscript. D.H.S. and L.L. conceived and designed the experiments. L.L. carried out chip design, fabrication, experiments and data analysis. M.K, S.B and X.L. conducted early-stage test experiments. I.M.M. carried out the material synthesis and obtained TEM images of the solution-phase samples. A.K. carried out part of the measurements. L.L., A.V.S., and D.H.S. wrote the manuscript. All authors have given approval to the final version of the manuscript. The authors declare no competing interests.

### Notes

The authors declare no competing financial interest.

### Acknowledgments

This work was supported by the National Science Foundation (CHE-2304936 to D.H.S.) and Welch Foundation (A-1547 to A.V.S). The nanofabrication was conducted in the AggieFab Nanofabrication Facility (RRID:SCR_023639) supported by the Texas A&M Engineering Experiment Station and Texas A&M University.

TOC

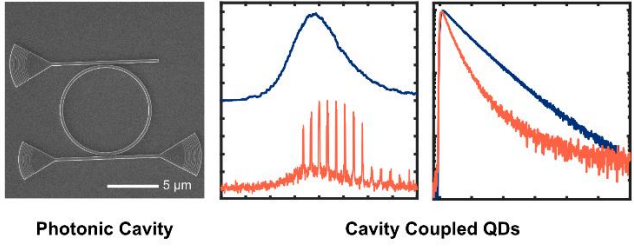

**Photonic Cavity**  **Cavity Coupled QDs**